# Identifying the regional drivers of influenza-like illness in Nova Scotia with dominance analysis

**Yigit Aydede**[#1] and **Jan Ditzen**[*]

[#]Saint Mary's University, Canada

[*]Free University of Bolzano, Italy

## Abstract

The spread of viral pathogens is inherently a spatial process. While the temporal aspects of viral spread at the epidemiological level have been increasingly well characterized, the spatial aspects of viral spread are still understudied due to a striking absence of theoretical expectations of how spatial dynamics may impact the temporal dynamics of viral populations. Characterizing the spatial transmission and understanding the factors driving it are important for anticipating local timing of disease incidence and for guiding more informed control strategies. Using a unique data set from Nova Scotia, the objective of this study is to apply a new novel method that recovers a spatial network of the influenza-like viral spread where the regions in their dominance are identified and ranked. We, then, focus on identifying regional predictors of those dominant regions.

## 1. Introduction

The recent advances in surveillance systems for infectious disease, capability of data collections and storage, and increased computational resources in the last decade have provided unprecedented tools for the scientific community to understand and, more importantly, combat the spread of infectious disease in populations. The importance of understanding the dynamics of underlying process in viral spread in response to its epidemiological factors such as weather-dependent correlates and most importantly non-pharmaceutical interventions has become more evident with the recent COVID-19 pandemic. While a large number of studies have examined individual-level risk factors for COVID-19, few studies have examined geographic hotspots and community drivers associated with spatial patterns in local transmission.

---

[1] Corresponding author (yigit.aydede@smu.ca).

The local public response to the COVID-19 pandemic in Nova Scotia provides a unique dataset similar to controlled clinical trials. Starting from the first week of March 2020, the local health authority established a set of rules for testing COVID-19. Only people with two or more symptoms of fever (higher than 38C), cough, sore throat, runny nose, and headache were asked to call 811, which served as a first assessment point run by registered nurses. The number of tests applied daily has dramatically increased in the range of 700-1200 per day with 1-3% of positivity rates. Since COVID-19 tests in the first four months of the first wave in 2020 are regulated by 811 Triage, each referral by the triage represents a significant and accurate level of information about the spatial and temporal distribution of influenza-like illness.

Studies investigating outbreaks by social geography provide us invaluable tools for understanding spatial and temporal determinants of the spread. Prior to the COVID-19 outbreak, it has been well-demonstrated that social, geographic, and economic factors impact the rate of infectious disease transmission [1,2,3,4,5,6]. Socio-spatial influences that have historically contributed to the rapid spread of infections are poor hygiene [6, 7, 8, 9, 10], low income [6,8,11,12 13,14], high population density [6,15], public and mass transit [16,17,18], malnutrition [19,20,21], and disadvantaged socioeconomic status [22]. There are several recent studies [23,24,25,26,27,28] focusing on spatial risk factors associated with the COVID-19 spread. The evidence unambiguously shows that climatic variables, mobility restrictions, and place-based factors like median household income, income inequality, and ethnic diversity in the local population explain a significant spatial variation in COVID-19 incidence. Overall, these studies report strong evidence of the spatial associations between the COVID-19 spread with selected local socioeconomic factors.

This study extends the previous work that investigates how contextual factors can contribute to the spatial distribution of a viral spread in several new directions. Unlike studies using cross-sectional incidence densities, regional hotspots, or spatial clusters (Spatial Scan Statistics [29]), we use a novel method related to a recent literature on granular time series [30] to explore the formation of spatial dependence in the network of regions. We identify and rank the dominance of each region in the spatial transmission network using the temporal dynamics in the data. With the application of this new method to epidemiological surveillance, we uncover "dominant regional drivers" and associated socio-spatial predictors rather than their associations with regional differences in incidence densities. The set of regional attributes used to identify socio-spatial factors of a spread in earlier studies are very limited due to its availability at a finer spatial scale (i.e., at the postal code). For example, most studies use county-level geographical classifications in the U.S. with few selected variables that are manually identified with a prior knowledge. We use regional data at the FSA (3-digit postal codes – Forward Sorting Area) level with more than 1400 demographic, economic, and social regional variables. Finally, to isolate important space-specific predictors of being a "dominant regional driver" of a viral pathogen, we exploit a framework that provides unbiased conditional variable importance yielded by random forest for feature selection.

To uncover spatial risk patterns for a viral spread, our study identifies the 18 regional drivers of influenza-like illness among 112 regions and significant space-specific characteristics associated with them. Those regional drivers are uniquely identified in terms of their community-level vulnerability, such as their demographic and economic characteristics. We recommend that predictive detection and spatial analysis be included in population-based surveillance strategies to better inform early case detection and prioritize healthcare resources.

**Figure 1 – COVID-19 Tests in Nova Scotia – (March-October 2020)**

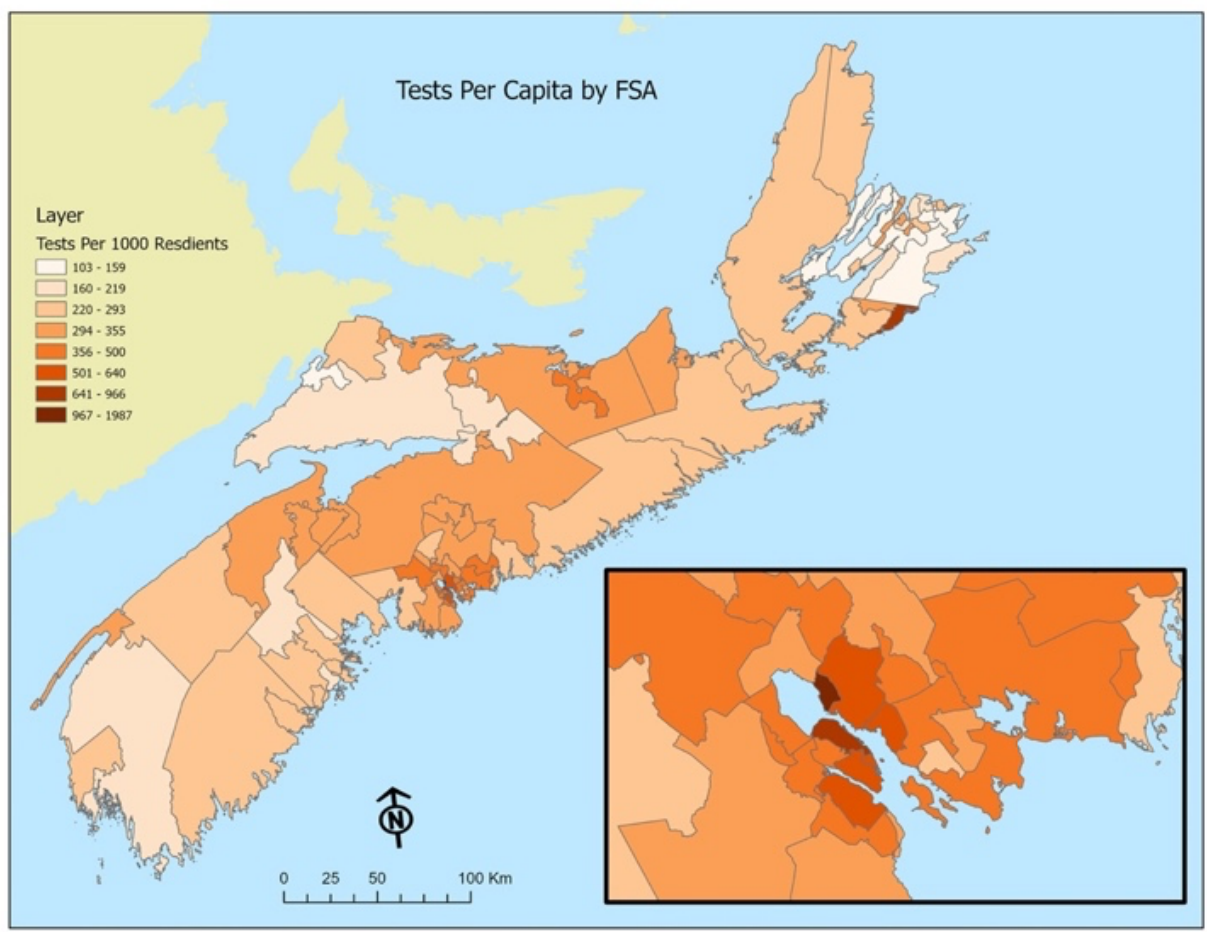

## 2. Data

This study uses a confidential and administrative dataset obtained as a part of Research Nova Scotia (RNS) project (HRC-2020-112) investigating the efficacy of mobility restrictions. The project's research team selected to the Nova Scotia COVID-19 Research Coalition is the only (as of March 2022) group of researchers that have been permitted to access the data out of Nova

Scotia Health Authority (NSHA). We start with a heatmap that summarizes the test numbers for Nova Scotia and Halifax Regional Municipality (HRM) in Figure 1.

Following the first case in February 2020, the local health authority established a set of rules for testing COVID-19:

1. People are asked to self-evaluate their symptoms whether they have two or more of the following symptoms: fever (higher than 38C), cough, sore throat, runny nose, and headache.
2. Those who meet the requirement are asked to call the 811 Triage.
3. A registered nurse assesses each caller and sends the caller to few local test centers.

We received the test data in May 2021 containing information by FSA with few unique characteristics: number of tests each day, average age, gender distribution, delay between test and test results, and the test results (positive or negative). Although the case numbers are very rare around 1000 incidents in the first three months of the pandemic, the test data during the first wave of COVID-19 provides rich information on the spatiotemporal distribution of influenza-like illness.

The data for the socio-spatial risk factors are obtained from Canadian Census Analyser (CHASS) for the year 2016 at the FSA level. The census profile variables are grouped in 16 subcategories: Population and Dwellings, Age and Sex, Dwelling (dwelling characteristics and household size), Marital Status, Language, Income, Knowledge of Language, Immigration, Aboriginals and Visible Minorities, Housing, Ethnic Origin, Education, Labour, Journey to Work, Language of Work, Mobility. In each category, variables represent averaged values at each FSA and for each gender type. When we include all categories, we obtain more than 1400 socio-spatial variables for each of 112 FSAs in Nova Scotia. Although the richness of data at this level of spatial scale is very desirable, it brings issues due to the curse of dimensionality, which will address later.

## 3. Methods

**Dominant units using graphical models**

Dominant units are units which influence the entire cross-section, that is all other units. In factor models they can often be modelled as observed common factors [30]. The degree a cross-sectional unit influences others varies. If a unit affects only the units closest to it, a shock of such unit will wear out when travelling through the network. This concept is called weak or spatial dependence and usually estimated by spatial methods. The figure on the left in Figure 2 describes such a setup. Dominant units in turn affect all other units and are at the core of a star type network (on the right): when the dominant unit Number 1 experiences a shock, this shock will influence all other units. Further, if the number of units increases, dominant units will affect those as well.

**Figure 2 – The difference between spatial and network analyses**

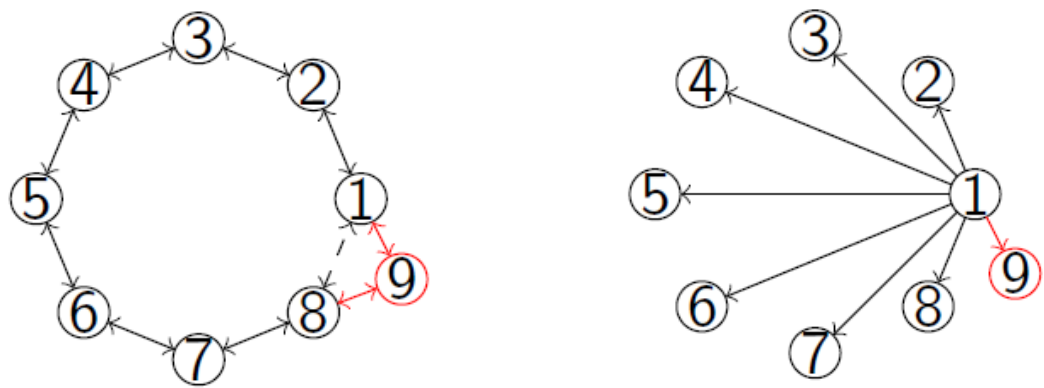

The identification of dominant units has recently received much attention and can be differentiated into two strands. One based on correlation matrices and the other on measures of connectedness. There are few recent examples for the former [30] and for the latter Kapetanios et al. [31] and Pesaran and Yang [32]. Other contributions such as Ditzen & Ravazzolo [33] or Gumundsson and Brownlees [34] combine both.

Kapetanios et al. [31] estimate the number of dominant units from residual variances from regressions of the individual time series on a prespecified number of common factors. In a second step the variances are thresholded to identify the dominant units. In case the dominant units affect only a subset of the other units, the authors propose a multiple testing approach [35] on the thresholded variances. A disadvantage of the approach is that it requires precise knowledge of the number of common factors and a threshold. Pesaran and Yang [32] propose an extremum estimator for degree of dominance based on a network or spatial approach. The estimator is based on the ratio between the maximum and of the sum of individual connectivity measures such as a spatial weight matrix.

Brownlees and Mesters [30] define dominant units with respect to the column norm of the concentration matrix. This implies that in a first step the inverse of the covariance matrix is calculated, limiting the approach to data with more observations than variables. The number of dominant units is then estimated using a growth criterion using a similar approach as the eigenvalue ratio criterion proposed in Ahn and Horenstein [36]. Finally, Gumundsson and Brownlees [34] identify dominant groups in a VAR using eigenvalues of the autoregressive coefficients. Both methods use the intuitive way to identify dominant units using covariance matrices. However, this is impossible if the number of variables is larger than the number of observations. A partial correlation matrix would be more precise, but hinges on estimation problems and suffers from potential noise in the correlations. A solution is to estimate the covariance matrix using dimensions reduction methods such as the lasso estimator. This approach is followed by Ditzen & Ravazzolo [33]. The authors suggest identifying dominant units using a two-step approach. In the first step a graphical network is estimated using a lasso estimator, following on the lines of Meinshausen and Bühlmann [37] and Sulaimanov and Koeppl [38]. In a second step the estimated inverse of the covariance matrix is then used to determine the number of factors as in Brownlees, & Mesters [30]. A disadvantage of the approach is that it assumes at least one dominant unit. Important to note is that, with exception of Gumundsson and Brownlees [34], the here mentioned literature originates from factor models. Dominant units are modelled as a special form of a common factor. This is very different from using lasso methods to select the appropriate number of lags in a VAR.

**Rigorous Lasso**

This paper follows the approach in Ditzen & Ravazzolo [33]. The authors suggest the following optimisation problem which is repeated for each cross-section:

$$\min_{\boldsymbol{\beta}_i} \frac{1}{T} \sum_{t=1}^{T} (x_{i,t} - x_{-i,t} \boldsymbol{\beta}_i')^2 + \frac{\lambda}{T} \sum_j^N \psi_j |\boldsymbol{\beta}_j|,$$

where $x_{i,t}$ is a T x 1 matrix containing the observations for the $i$-th unit. $x_{-i,t}$ is a T x (N-1) matrix containing all other cross-sections. $\beta_i$ is a 1 x (N-1) sparse vector containing past lasso OLS estimates. Non-zero elements in $\kappa_i$ imply that the respective cross-section influences cross-section $i$. In a graphical setting, the $\beta_{ij} \neq 0$ implies that unit $i$ and $j$ are connected and thus they have an edge connecting them. $\frac{\lambda}{T} \sum_j^N \psi_j |\beta_j|$ is the penalty term with the tuning parameter $\lambda$ and the loading $\psi_j$. Both need to be specified prior to estimation, depending on the estimation method. Ditzen & Ravazzolo (2022) find that the rigorous (or plugin) lasso [39,40,41] or the adaptive lasso [42,43] works best to uncover the graphical representation in a framework with dominant units. The rigorous lasso has the advantage that it is data driven and therefore does not require any parametrisation or specification of hyperparameters. In detail, it sets $\lambda = 2c \sqrt{n} \phi^{-1}\left(1 - \frac{\gamma}{2p}\right)$ with $c$ a slack parameter, $\gamma$ the probability of the regularisation event [40]. Both are commonly set to $c = 1.1$, $\gamma = 0.1/\log(n)$. The penalty loading $\psi_j$ is estimated as $\widehat{\psi}_j^2 = \frac{1}{NT} \sum_{i=1}^{N} \left(\sum_{t=1}^{T} \ddot{x}_{i,t,j} \widehat{\epsilon_{i,t}}\right)^2$ with $\ddot{x}_{i,t,j}$ as the deviations of the unit specific means and $\widehat{\epsilon_{i,t}}$ is obtained from an auxiliary regression. Ahrens et. al [41] show that $\widehat{\psi}_j^2$ using autocorrelation and autocorrelation heteroskedasticity robust estimators in the presence of autocorrelation and heteroskedastic errors.

If the adaptive lasso is used, the penalty loadings $\psi_j$ are estimated using an unbiased and consistent estimator. Depending on the ratio of variables to observations uni- or multivariate OLS is proven to lead to the desired oracle properties [42,44]. The tuning parameter is commonly specified by either cross-validation or information criteria such as the AIC or BIC. As the rigorous and adaptive lasso both perform similar with respect the identification of the dominant units [33], we choose the data driven rigorous lasso for our analysis.

The estimated $\hat{\beta}_i$ are then stacked together $\hat{\beta} = (\hat{\beta}_1, \hat{\beta}_2, \ldots, \hat{\beta}_N)'$ into a N x N matrix, where the diagonal elements are zero:

$$\boldsymbol{\beta} = (\boldsymbol{\beta}_1, \ldots, \boldsymbol{\beta}_{77}) = \begin{pmatrix} 0 & \beta_{1,1} & \cdots & \cdots & \beta_{1,77} \\ \beta_{2,1} & 0 & \beta_{2,3} & \cdots & \beta_{2,77} \\ \vdots & \vdots & \vdots & \vdots & \vdots \\ \vdots & \vdots & \vdots & \vdots & \vdots \\ \beta_{77,1} & \cdots & \cdots & \cdots & 0 \end{pmatrix}$$

Based on Sulaimanov and Koeppl [38] we obtain the concentration matrix by multiply the estimated coefficients with the inverse of the unit specific residual variances:

$$\hat{\kappa} = \widehat{D}(I - \hat{\beta})$$
$$\widehat{D} = diag(\hat{\sigma}_1^{-2}, \ldots, \hat{\sigma}_{77}^{-2})$$

The idea behind the step is that the sparsity of the estimated coefficients carries over to the concentration matrix.

Following Brownlees & Mesters [30] and Ditzen & Ravazzolo [33] the column norm is used to identify dominant units. Define $\tilde{\kappa}_i$ as the $i$-th column of $\hat{\kappa}$, then unit $i$ dominates unit $j$ if the column norm of $i$ is larger than of $j$, hence $||\tilde{\kappa}_i|| > ||\tilde{\kappa}_j||$. This implies that a shock to unit $i$ has a larger effect to the entire network than a shock to unit $j$. Brownlees & Mesters [30] and Ditzen & Ravazzolo [33] suggest identifying the number of dominant units by ordering the column norms by their size and then calculate the growth rate. The maximum of the growth rate defines the number of dominant units $k$:

$$k = \max_{i=1,\ldots,N} ||\tilde{\kappa}_i||/||\tilde{\kappa}_{i+1}||$$

We use the incidence density (or incidence rate, which is the number of symptomatic people/population) between the 1$^{st}$ of March and the 1$^{st}$ of July in 2020. After standardizing the data by taking the first-difference and scaling, we identify 18 driver regions out of 77 FSA's as shown in Figure 3.

**Figure 3: Regional drivers of influenza-like illness in Nova Scotia - 2020**

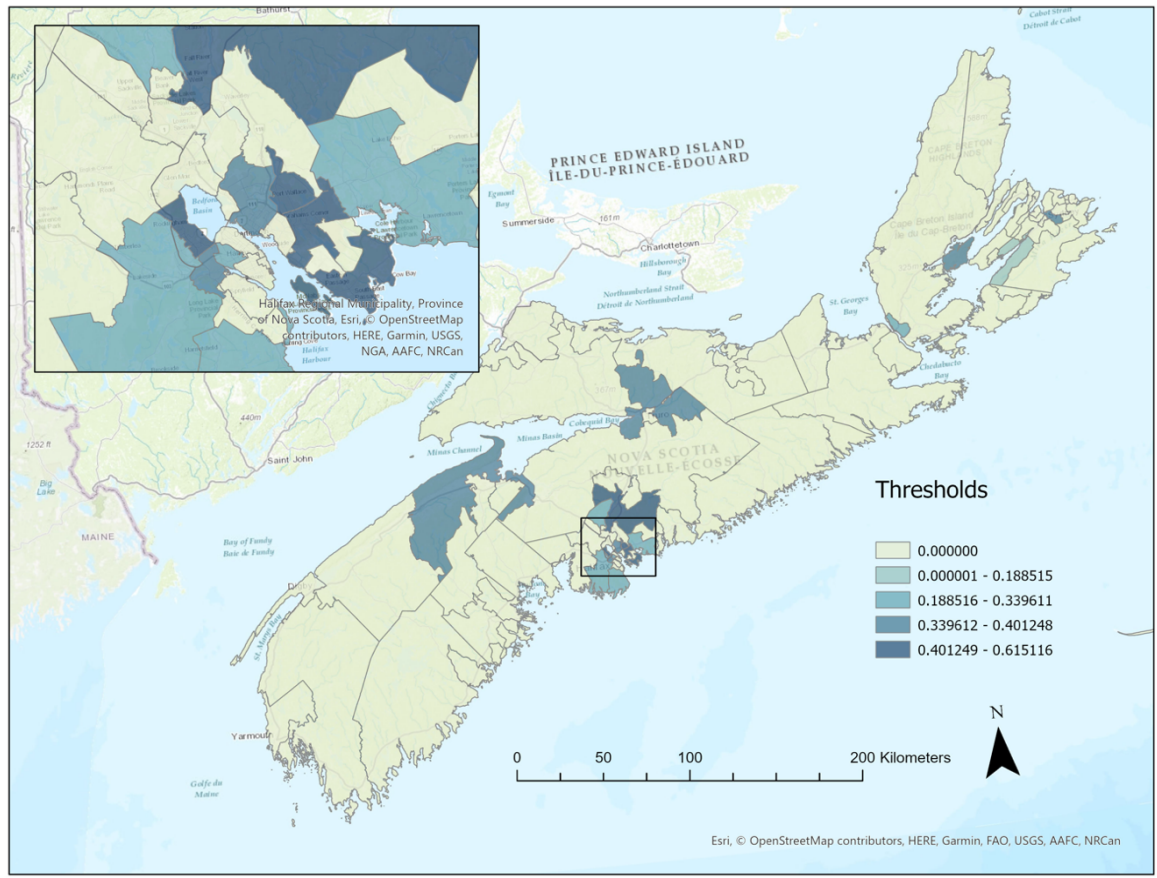

It could be thought that column norms and their order might be parallel with the level of "infection density" in each FSA. In other words, one would think that the magnitude of infection density in each FSA would help identify the regional spreaders similar to the process of our dominant unit analysis. The correlation between column norms selected by rigorous lasso and infection densities in each dominant FSA is 0.31. The R-squared in the regression between norms and densities is 0.0807 with a statistically insignificant coefficient, which is 0.328 with the standard error of 0.22. This does not change with a logarithmic transformation nor polynomial terms.

This shows that a region with a low infection density could be a regional driver while a hotspot with a higher density could be a submissive region in a viral spread. The important distinction in recognizing the regional drivers is that the dominant unit analysis encompasses the temporal dynamics of the viral spread in its network structure, while the infections densities are just cumulative sums reflecting only cross-sectional differences.

There are few other studies to characterize spatiotemporal trends in the spread of a viral pathogen. The literature is mostly focused on how much impact the influenza-like illness of one place could have on its neighbours [45,46] by utilizing personal contact data. To overcome the data requirements, Qiu et al. [47] proposed a new concept called spatiotemporal route to display the potential transmission directions and the sizes of those transmission effects. This concept was defined as the time-lagged association among influenza surveillance data across different places using vector-autoregressive models (VAR). Later, the same method is used by Wang et al [48] to forecast the spread of the viral spread. Although VAR models are good for forecasting in complex spatiotemporal networks, they are not well-designed to reveal the structure of complex networks specially for understanding the "dominant units" in panel data. While the machine learning assisted approach to recover the underlying network in Qiu et. al. [47] is similar to ours, there are key differences between the works. The main difference is that we identify which units are dominant. Qiu et. al. [47] treat all non-zero connections as channels to spread influenza, independent of their actual importance. Further, Qiu et. al. [47] try to identify the spread of influenza increasing sequentially the lag length and each lag length has its own network. Therefore, the spread is dependent on time, while in our approach the spread is time independent.

Next, the task is to see what regional characteristics make those eighteen regional drivers different than the rest of the regions.

## 4. Results

**Random Forests**

High-dimensional data have been a subject of theoretical and applied investigations in the last few decades sparked by advances in technology that enable to process situations where the number of covariates or predictors is much larger than the number of data points (i.e., $n < p$). As there is often a sense that many of the features are of minor importance in the empirical literature of different domains, researchers select the number of explanatory variables by hand, rather than choosing them in a data-dependent manner. Yet, we may not know ex ante which of the features matter and which can be dropped from the analysis without substantially hurting the predictive power.

With the advancement of machine learning, identifying relevant predictor variables, rather than only predicting the response, becomes a major motivation in many applications. Tree-based methods can help identify relevant predictors even in a high dimensional setting involving complex interactions. Random forests [49] algorithm produces many single trees based on randomly selected a subset of observations and features. Since the algorithm leads to many single trees (like a forest) with a sufficient variation, the averaging them provides relatively a smooth and, more importantly, better predictive power than a single tree. Random forests and regression trees are particularly effective in settings with a large number of features that are not related to the outcome, that is, settings with sparsity [50]. The splits will generally ignore those covariates, and as a result, the performance will remain strong even in settings with many features [51,52]. Variable importance measures for random forests, for example, have been a major tool as a means of variable selection in many classification tasks in bioinformatics and genetic epidemiology including large-scale association studies for complex genetic diseases [53,54]. A comparison of random forests and other classification methods for the analysis of gene expression data is presented by Díaz-Uriarte and Alvarez de Andrés [55], who propose a new gene selection method based on random forests for sample classification with microarray data.

After removing four FSA's with missing observations, we have 77 FSA's and 1378 regional predictors. We obtain the results reported below based on 2000 runs of the random forest algorithm. This is because a random forest algorithm gets its final estimations as the average of trees using bootstrapped subsamples of observations and features. Therefore, two random forest model estimated from the same data may have slightly different results due the fact that its trees would be different in each model. This particularly true when we have the data that has a very low n/p ratio, which is around 5.0% (77/1378) in our case.

Although our main task is to identify the major predictors of regional spreaders, since the reliability of their selection depends on predictive capacity of the entire model, we first start with the exploring the model performance.

**Model performance**

Achieving our objective – finding the most important predictors – is contingent on the model's predictive performance. In explanatory modelling where obtaining the causal relationship is the main purpose, we try to see how well the model's predictions explain (fit) outcomes in the data used for developing the model. In our case, however, the relevant question is how well the model predicts the outcome for new observations. With the bootstrap resampling process for each tree, random forests have an efficient and reasonable approximation of the test error calculated from out-of-bag (OOB) sets. When bootstrap aggregating is performed, two independent sets are created. One set, the bootstrap sample, is the data chosen to be "in-the-bag" by sampling with replacement. The out-of-bag set is all data not chosen in the sampling process. Hence, there is no need for cross-validation or a separate test set to obtain an unbiased estimate of the prediction error. It is estimated internally as each tree is constructed using a different bootstrap sample from the original data. About one-third of the cases (observations) are left out of the bootstrap sample and not used in the construction of the $k^{th}$ tree. In this way, a test set classification is obtained for each case in about one-third of the trees. In each run, the class is selected when it gets most of the votes among the OOB observations. The proportion of times that the selected class for the observation is not equal to the true class over all observations in OOB set is called the OOB error estimate. This has proven to be unbiased in many tests [56]. Thus, instead of using a conventional 3-layer (Training -Validation-Test sets) cross validation process, we use OOB test errors of 1500 trees in each random forest algorithm averaged after 2000 runs. Note that the forest's variance decreases as the number of trees grows. Thus, more accurate predictions are likely to be obtained by choosing a large number of trees. In our random forest algorithm, setting the number of trees at 1500 is the optimal in terms of prediction accuracy and computational efficiency.

After eliminating a few rural FSAs with a very low population density and one outlier region, we include 74 total FSAs (18 "Dominants", 56 "Followers") in our initial application. Although the class balance is not skewed drastically ($18/56 = 0.321$), we use a stratified random forest algorithm to reduce the bias in each split. We also use the default setting of randomly selected features ("mtry" $= \sqrt{p} = 37$) in each tree. The OOB results are shown in Table 1. Note that the confusion table is obtained by averaging 2000 runs and the values rounded down to the nearest integer. All the results indicate a decent prediction accuracy of our initial application without preprocessing and explicit training.

**Table 1: OOB Results of 2000 runs**

| Predicted | Actual Dominants | Actual Followers | | 95% CI | |
|---|---|---|---|---|---|
| **Dominants** | 13 | 15 | | | |
| **Followers** | 5 | 41 | | | |
| OOB error rate | 33.78% | | | | |
| **PPV** | | | 0.46428571 | 0.45428971 | 0.47428171 |
| **NPV** | | | 0.89130435 | 0.88130835 | 0.90247635 |
| **Specificity - TNR** | | | 0.89130435 | 0.88130835 | 0.90228035 |
| **Sensitivity - TPR** | | | 0.72222222 | 0.71222622 | 0.73221822 |
| **Balanced Accuracy** | | | 0.80676329 | 0.79676729 | 0.81724929 |

Note: PPV-Positive predictive value, NPV-Negative predictive value, TNR – True negative rate, TPR – True positive rate.

Our outcome variable is categorical ($Y = 1$ spreader, 0 otherwise). Random forest algorithm calculates the predicted probability of success ($Y = 1$). If the probability is larger than a fixed cut-off threshold (c), then we assume that the model predicts success ($Y = 1$); otherwise, we assume that it predicts failure. As a result of such a procedure, the comparison of the observed and predicted values of the dependent variable summarized in a confusion table depends on the cut-off threshold (c). The predictive accuracy of a model as a function of threshold can be summarized by Area Under Curve (AUC) of Receiver Operating Characteristics (ROC). The ROC curve indicates a trade-off between True Positive Rate (TPR) and False Positive Rate (FPR). Hence, the success of a model comes with its predictions that increases TPR without raising FPR. Using the predicted OOB probabilities averaged over 1500 trees, we find the optimal cut-off threshold (c) and calculate AUC for each run. After taking its average over 2000 runs, we find AUC 74.6% with a standard error of 0.019051.

Although random forest algorithms do not need an explicit training, the number of selected features ($\sqrt{p}$) can be set as a hyperparameter and identified by a grid-search. We set a 5-fold cross-validation process repeated 5 times searching a grid of randomly selected number of features in each run. With this process, the OOB AUC improves to 81.56% with a standard error of 0.03787. These results report the out-of-sample prediction accuracy. When we apply the trained model for in-sample predictions using the entire data, AUC improves to 89.11% indicating a good internal validity of the algorithm. Finally, we also implement some preprocessing operations to reduce the number of highly correlated predictors as well as near-zero-variance features. Although the results after this preprocessing do not significantly improve, the selection of important predictors would benefit from this process, which is what we are going to discuss next.

**Variable importance**

There are several options to evaluate how important is the variable $x$ in predictions. We use a permutation-based variable-importance in which the effect of a variable is removed through a random reshuffling of the data in $x$. This method takes the original data under $x$, permutates (mixes) its values, and gets "new" data, on which computes the weighted decrease of impurity corresponding to splits along the variable $x$ and averages this quantity over all trees. If a variable is an important predictor in the model, after its permutation, the mean decrease impurity (MDI) rises. It is shown by Louppe et al. [57] that building a tree with additional irrelevant variables does not alter the importance of relevant variables in an infinite sample setting. Another measure of significance, Mean Decrease Accuracy (MDA), stems from the idea that if the variable is not important, rearranging its values should not degrade prediction accuracy. The MDA relies on a different principle and uses the out-of-bag error estimate.

The empirical properties of the MDI criterion have been extensively explored and compared in the statistical computing literature. For example, Archer and Kimes [58] point out that MDA (and MDI) may behave poorly when correlation increases, which is experimentally tested by Auret and Aldrich [59] and Tolosi and Lengauer [60]. Auret and Aldrich [59] show three trends in their simulation study: (1) as the association between a variable and the response increases, the proportion of times that that variable is correctly identified increases; (2) as the correlation of within-group variables increases, the proportion of times the true variable is identified decreases; (3) as the correlation of within-group variables increases, the proportion of times the true group is identified increases. The problem with correlated predictors can be explained as follows. A variable $x_i$ is permuted to destroy the association between $x_i$ and the response, $y$. However, this permutation also destroys the association between $x_i$ and the other input variables $x_{j(j \neq i)}$. Thus, a decrease in model accuracy can then imply dependence of $y$ on $x_i$ or dependence of $x_i$ on any of $x_{j(j \neq i)}$, the latter not being useful in the context of variable importance measures.

Strobl et al. [61] and Hothorn et al. [62] have suggested a conditional permutation framework to reduce this effect and applied a conditional permutation to their new random forest algorithm forests (Conditional Forest – CF). Although CF appears to work better for identification of significant variables, it has its own shortcomings specially when the number of trees is large [59,63]. Hence, it is suggested that, when it is feasible, the best practice is to include expert knowledge of the process under consideration in data preprocessing by reducing the presence of correlated variables. We implement a two-step preprocessing: first, we removed variables that have a very low variance. The concern here is that when the data are split into bootstrap samples, a few samples may have an undue influence on the model. Second, we try to identify and reduce the level of within-group correlations between the predictors to address the concern related to a possible bias in MDI and MDA. For the first step, we removed 126 variables that have a near-zero variance. In the second step, we calculated mean absolute pair-wise correlation for each variable. We look at their within-group and across-group correlations. After setting up the cut-off correlation coefficient at 0.85, we manually checked those removed variables to make sure that their respected groups would not dismiss important predictors. We removed 679 additional variables. After these restrictions, we include 375 regional predictors in our algorithms. Although the level of elimination among predictors may seem significant, the census file follows a hierarchical aggregation for each group in total, male, and female splits. Hence, many purged predictors have either in top level aggregations or related to gender partitions. We also tried several different cut-off points ranging from 0.85 to 0.99 to see the union of top 30 predictors across all models. Although their order changes, more than 60% of these top predictors remains in all cut-off points. Moreover, the predictive accuracy remains about the same validating that the removed variables are accurately selected.

The results for the top 30 predictors are shown in Figure 4. The first figure shows the top predictors ranked by the frequency of how many times each predictor is selected in 2000 runs by MDI. The bottom plot shows the top predictors ranked by their MDI averaged over 2000 runs. The list of predictors in both plots are the same.

**Figure 4 – Variable Importance measures over 2000 Random Forests**

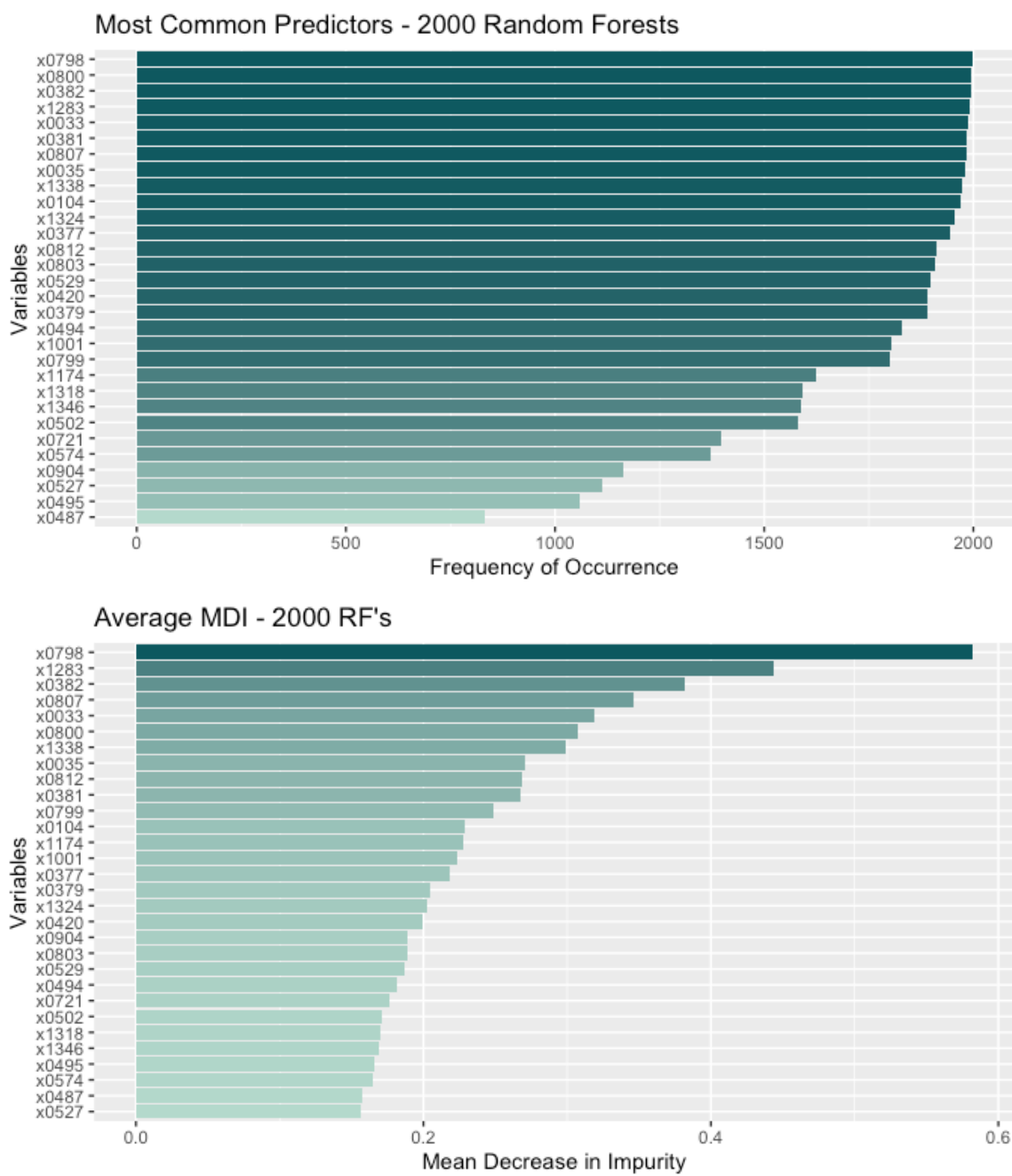

Note: The names of variable codes are given in Table 2.

Variable importance applications provide global measures not local or directional ones. Directional local variable importance measures based on ceteris paribus profiles could be obtained by partial dependence profiles at each prediction point, that is for each major FSA's. There are several tools for instance-level explorations and for understanding local-level profiles [64].

Table 2: Differences-in-means comparisons for selected predictors

| Code | MDI | Followers | Dominants | Names |
|---|---|---|---|---|
| x0798 | 0.582 | 50.6 | 61.2 | Homeowners with a mortgage (%) |
| x1283 | 0.444 | 40.1 | 42.7 | Weeks worked |
| x0382 | 0.382 | 64.2 | 70.5 | With employment income (%) |
| x0807 | 0.346 | 764.9 | 901.8 | Median monthly rent $ |
| x0033 | 0.319 | 64.8 | 67.8 | People between 15-65 (%) |
| x0800 | 0.308 | 793.9 | 1,080.4 | Med. monthly shelter cost – owners |
| x1338 | 0.299 | 218.2 | 727.5 | Using public transit to work |
| x0035 | 0.271 | 2.4 | 1.7 | People 65+ (%) |
| x0812 | 0.268 | 20.9 | 36.9 | People with Inuit origin |
| x0381 | 0.268 | 81.6 | 87.1 | Total market income $ |
| x0799 | 0.249 | 11.6 | 12.0 | Spent 30% + inc. on shelter (%) |
| x0104 | 0.228 | 211.0 | 146.7 | People speak French at home |
| x1174 | 0.228 | 1.7 | 7.2 | People in military service |
| x1001 | 0.223 | 4.6 | 17.2 | People with African origin |
| x0377 | 0.218 | 8,111.6 | 5,910.1 | Median government transfers $ |
| x0379 | 0.205 | 28,918.8 | 33,548.6 | Median employment income $ |
| x1324 | 0.202 | 426.0 | 904.7 | People working in public sector |
| x0420 | 0.199 | 66.0 | 70.8 | With employment income (%) |
| x0904 | 0.189 | 2.9 | 11.4 | People with Caribbean origin |
| x0803 | 0.189 | 218,969.0 | 267,140.6 | Average value of dwelling $ |
| x0529 | 0.187 | 17.6 | 12.9 | People 65+ with low-income |
| x0494 | 0.181 | 88,184.2 | 91,594.5 | Med. income - families with kids $ |
| x0721 | 0.176 | 13.8 | 26.1 | People with aboriginal ancestries |
| x0502 | 0.171 | 26,360.5 | 29,776.0 | Med. income - noneconomic fam. $ |
| x1318 | 0.170 | 254.2 | 444.4 | People in waste management ind. |
| x1346 | 0.168 | 298.6 | 424.2 | People travel to work 45-59 min. |
| x0495 | 0.166 | 3.8 | 3.9 | Average family size - with kids |
| x0574 | 0.165 | 14.4 | 34.2 | Immigrants from Americas (other) |
| x0487 | 0.158 | 2.7 | 2.8 | Average family size - Overall |
| x0527 | 0.156 | 27.1 | 21.9 | People with low-income 0-17 |

For this study, however, we apply a simple difference-in-means comparison for selected predictors as reported in Table 2 for two reasons. First, we do not have a specific objective to discover causal relationships. This would be achieved better with semi-parametric applications like MARS (multiple adaptive regression splines). Second, extracting local variable importance is an active research area and most of the tools developed for understanding local and directional effects for random forests rely on strong assumptions such as independence between predictors.

The results in Table 2 can be summarized as follows: the driver regions are associated with high-income populations who are at their prime age (15-65), employed, and working on developing their housing wealth. These regions have also a profile associated with more public workers and military families. People in spreader regions use public transportations and travel to work more than 45 minutes, which is about three times as much as other regions. In terms of demographics, blacks, Inuit's, and people with Caribbean and Latin America origins are more populated in driver regions. However, the different cut-off correlations result different ethnic origins, which is perhaps due to a high correlation between different ethnic groups. To solve this problem, we included one major group, immigration density, that is selected in the top ten predictors in all cut-off points. Our findings are consistent with the evidence that shows that mobility restrictions, median household income, income inequality, and ethnic diversity in the local population explain a significant spatial variation in COVID-19 incidence. Finally, we should note that these interpretations are based on the limited capacity of differences-in-means in Table 2 and cannot properly reflect possible effects of complex interactions between the predictors.

## 6. Concluding Remarks

Studies investigating outbreaks by social geography provide us invaluable tools for understanding spatial and temporal determinants of the spread. Prior to the COVID-19 outbreak, it has been well-demonstrated that social, geographic, and economic factors impact the rate of infectious disease transmission. This study extends the previous work that investigates how contextual factors can contribute to the spatial distribution of a viral spread in several new directions.

The local public response to the COVID-19 pandemic in Nova Scotia provides a unique dataset similar to controlled clinical trials. We use a novel method related to a recent literature on granular time series to explore the formation of spatial dependence in the network of regions. We identify and rank the dominance of each region in the spatial transmission network using the temporal dynamics in the data. With the application of this new method to epidemiological surveillance, we uncover "dominant regional drivers" and associated socio-spatial predictors. To uncover spatial risk patterns for a viral spread, our study identifies the 18 dominant regional drivers among 112 regions and significant space-specific characteristics associated with them. Those regional drivers are uniquely identified in terms of their community-level vulnerability, such as their demographic and economic characteristics. We recommend that predictive detection and spatial analysis be included in population-based surveillance strategies to better inform early case detection and prioritize healthcare resources.

# References

1. Oestergaard LB, Schmiegelow MD, Bruun NE, Skov RL, Petersen A, Andersen PS, Torp-Pedersen C. The associations between socioeconomic status and risk of Staphylococcus aureus bacteremia and subsequent endocarditis - a Danish nationwide cohort study. *BMC Infect Dis.* 17(1). 1-9 (2017).
2. Gares V, Panico L, Castagne R, Delpierre C, Kelly-Irving M. The role of the early social environment on Epstein Barr virus infection: a prospective observational design using the Millennium Cohort Study. *Epidemiol Infect.* 145(16), 3405-3412 (2017).
3. Doherty, I.A., Leone, P.A., Aral, S.O. Social determinants of HIV infection in the deep south. *Am. J. Public Health* 97 (3), 391 -391 (2007).
4. Coffey, P.M., Ralph, A.P., Krause, V.L. The role of social determinants of health in the risk and prevention of group A streptococcal infection, acute rheumatic fever and rheumatic heart disease: a systematic review. *PLoS Negl. Trop. Dis.* 12 (6), 42-54 (2018).
5. Rosenthal, J. Climate change and the geographic distribution of infectious diseases. *Ecohealth* 6 (4), 489–495 (2009).
6. McMichael, A.J. Environmental and social influences on emerging infectious diseases: past, present and future. *Philos. Trans. R. Soc. Lond. B Biol. Sci.* 359 (1447), 1049–1058 (2004).
7. Morse, S.S. Factors in the emergence of infectious diseases. *Emerg. Infect. Dis.* 1 (1), 7–15 (1995).
8. Rao, G.G. Risk factors for the spread of antibiotic-resistant bacteria. *Drugs* 55 (3), 323–330 (1998).
9. Bula-Rudas, F.J., Rathore, M.H., Maraqa, N.F. Salmonella infections in childhood. Adv. *Pediatr.* 62 (1), 29–58 (2015).
10. Hall, C.B. The spread of influenza and other respiratory viruses: complexities and conjectures. *Clin. Infect. Dis.* 45 (3), 353–359 (2007).
11. Relman, D.A., Choffnes, E.R. Institute of Medicine (US) Forum on Microbial Threats. *The Causes and Impacts of Neglected Tropical and Zoonotic Diseases: Opportunities for Integrated Intervention Strategies.* Washington (DC): National Academies Press (US) (2011).
12. World Health Organization & UNICEF/UNDP/World Bank/WHO Special Programme for Research and Training in Tropical Diseases. (2012). *Global report for research on infectious diseases of poverty 2012.* World Health Organization. https://apps.who.int/iris/handle/10665/44850.
13. Farmer, P. Social inequalities and emerging infectious diseases. *Emerg. Infect. Dis.* 2 (4), 259–269 (1996).
14. Bonds, M.H., Dobson, A.P., Keenan, D.C. Disease ecology, biodiversity, and the latitudinal gradient in income. *PLoS Biol.* 10 (12) (2012).
15. Weiss, R.A., McMichael, A.J. Social and environmental risk factors in the emergence of infectious diseases. *Nat. Med.* 10 (12), S70–S76 (2004).

16. Morse, S.S. Factors in the emergence of infectious diseases. *Emerg. Infect. Dis.* 1 (1), 7–15 (1995).
17. Goscé, L., Johansson, A. Analysing the link between public transport use and airborne transmission: mobility and contagion in the London underground. *Environ. Health* 17 (1), 84. (2018).
18. Nasir, Z.A., Campos, L.C., Christie, N., Colbeck, I. Airborne biological hazards and urban transport infrastructure: current challenges and future directions. *Environ Sci Pollut Res Int.* 23(15), 15757-66 (2016).
19. Lederberg, Joshua, Hamburg Margaret, A., Smolinski, Mark S., Institute of Medicine (US) Committee on Emerging Microbial Threats to Health in the 21st Century. *Microbial Threats to Health: Emergence, Detection, and Response.* Washington (DC): National Academies Press (US); 2003. PMID: 25057653.
20. Schaible, U.E., Stefan, H.E. Malnutrition and infection: complex mechanisms and global impacts. PLoS Med. 4(5), e115. (2007).
21. Peter Katona, Judit Katona-Apte, The Interaction between Nutrition and Infection, *Clinical Infectious Diseases*, 46(10), 1582–1588 (2008).
22. Khalatbari-Soltani S, Cumming RC, Delpierre C, Kelly-Irving M. Importance of collecting data on socioeconomic determinants from the early stage of the COVID-19 outbreak onwards. *J Epidemiol Community Health.* 74(8), 620-623 (2020).
23. Franch-Pardo, I., Napoletano, BM, Rosete-Verges, F, Billa, L. Spatial analysis and GIS in the study of COVID-19. A review, *Science of The Total Environment*, 739 (2020).
24. Andersen LM, Harden SR, Sugg MM PhD, Runkle JD PhD, Lundquist TE. Analyzing the spatial determinants of local Covid-19 transmission in the United States. *Sci Total Environ.* (2021) Feb 1; 754:142396.
25. Mollalo, A., Vahedi, B., Rivera KM., GIS-based spatial modelling of COVID-19 incidence rate in the continental United States, Science of The Total Environment, 728 (2020).
26. Henning A, McLaughlin C, Armen S, Allen S. Socio-spatial influences on the prevalence of COVID-19 in central Pennsylvania. *Spat Spatiotemporal Epidemiol.* 37 (2021).
27. Wu, Xiao, et al., Exposure to air pollution and COVID-19 mortality in the United States. *Science Advances*, 6(45) (2020).
28. Kamel Boulos MN, Geraghty EM. Geographical tracking and mapping of coronavirus disease COVID-19/severe acute respiratory syndrome coronavirus 2 (SARS-CoV-2) epidemic and associated events around the world: how 21st century GIS technologies are supporting the global fight against outbreaks and epidemics. *Int J Health Geogr.* 19(1) (2020).
29. Otani, T., & Takahashi, K. Flexible Scan Statistics for Detecting Spatial Disease Clusters: The rflexscan R Package. *Journal of Statistical Software*, 99(13), 1–29 (2021).
30. Brownlees, C., & Mesters, G. Detecting granular time series in large panels. *Journal of Econometrics*, *220*(2), 544–561 (2021).
31. Kapetanios, G., Pesaran, M. H., & Reese, S. Detection of units with pervasive effects in large panel data models. *Journal of Econometrics.* 221(2), 510-541 (2021).

32. Pesaran, M. H., & Yang, C. F. Econometric analysis of production networks with dominant units. *Journal of Econometrics*, *219*(2), 507–541 (2020).
33. Ditzen, J., & Ravazzolo, F. Dominant Drivers of National Inflation. *arXiv:2212.05841[econ.EM]* https://doi.org/10.48550/arXiv.2212.05841 (2022).
34. Guðmundsson, G. S., & Brownlees, C. Detecting groups in large vector autoregressions. *Journal of Econometrics*, 225(1), 2–26 (2021).
35. Bailey, N., Pesaran, M. H., & Smith, L. V. A multiple testing approach to the regularisation of large sample correlation matrices. *Journal of Econometrics*, *208*(2), 507–534 (2019).
36. Ahn, Seung C. and Alex R. Horenstein. Eigenvalue Ratio Test for the Number of Factors. *Econometrica* 81(3):1203–1227 (2013).
37. Meinshausen, N., & Bühlmann, P. High-dimensional graphs and variable selection with the Lasso. *Annals of Statistics*, *34*(3), 1436–1462 (2006).
38. Sulaimanov, N., & Koeppl, H. Graph reconstruction using covariance-based methods. *Eurasip Journal on Bioinformatics and Systems Biology*, *2016*(1).
39. Bickel, P. J., Ritov, Y., & Tsybakov, A. B. Simultaneous analysis of lasso and dantzig selector. Annals of Statistics, *37*(4), 1705–1732 (2009).
40. Belloni, A., Chernozhukov, V., Hansen, C., & Kozbur, D. Inference in High-Dimensional Panel Models with an Application to Gun Control. *Journal of Business & Economic Statistics*. 115 (2016).
41. Ahrens, A., Aitken, C., Ditzen, J., Ersoy, E., Kohns, D., & Schaffer, M. E. A Theory-Based Lasso for Time-Series Data. In N. N. Thach, V. Kreinovich, & N. D. Trung (Eds.), *Data Science for Financial Econometrics. Studies in Computational Intelligence, vol 898.* (898th ed., pp. 3–36).
42. Zou, H. The adaptive lasso and its oracle properties. *Journal of the American Statistical Association*, *101*(476), 1418–1429 (2006).
43. Medeiros, M. C., & Mendes, E. F. l1-regularization of high-dimensional time-series models with non-Gaussian and heteroskedastic errors. *Journal of Econometrics*, *191*(1), 255–271 (2016).
44. Huang, J., Ma, S., & Zhang, C.H. Adaptive Lasso for Sparse High-Dimensional Regression Models Supplement. *Statistica Sinica* 18, 1603–1618 (2008).
45. Fu X, Small M, Chen G. *Propagation dynamics on complex networks.* 1st ed. West Sussex: Wiley; 2014.
46. Pei S, Kandula S, Yang W, Shaman J. Forecasting the spatial transmission of influenza in the United States. *Proc Natl Acad Sci USA.* 115(11), 2752–7 (2018).
47. Qiu, J., Wang, H., Hu, L. et al. Spatial transmission network construction of influenza-like illness using dynamic Bayesian network and vector-autoregressive moving average model. *BMC Infect Dis* 21, 164 (2021).
48. Wang H, Qiu J, Li C, Wan H, Yang C and Zhang T. Applying the Spatial Transmission Network to the Forecast of Infectious Diseases Across Multiple Regions. *Front. Public Health* 10 (2022).

49. Breiman, L. Random Forests. *Machine Learning* 45, 5–32 (2001).
50. Athey, S. and Imbens, GW. Machine Learning Methods that Economists should know about. Annual Review of Economics 11(1) 685-725 (2019).
51. Wager S, Athey S. Estimation and inference of heterogeneous treatment effects using random forests. *J. Am. Stat. Assoc.* 113, 1228–42 (2017).
52. Biau, G., Scornet, E. A random forest guided tour. *TEST* 25, 197–227 (2016).
53. Lunetta KL, Hayward LB, Segal J, Eerdewegh PV: Screening Large- Scale Association Study Data: Exploiting Interactions Using Random Forests. *BMC Genetics* 5:32 (2004).
54. Bureau A, Dupuis J, Falls K, Lunetta KL, Hayward B, Keith TP, Eerdewegh PV: Identifying SNPs Predictive of Phenotype Using Random Forests. *Genetic Epidemiology*, 28(2), 171-182 (2005).
55. Díaz-Uriarte, R. and Alvarez de Andrés, S. Gene selection and classification of microarray data using random forest. *BMC Bioinformatics*, 7, 3 (2006).
56. Liaw, A. and Wiener, M. Classification and Regression by Randomforest. *R News*, 2, 18-22 (2002).
57. Louppe, G., Wehenkel, L., Sutera, A., Geurts, P. Understanding variable importances in Forests of randomized trees. *Advances in Neural Information Processing Systems.* 26. (2013).
58. Archer, K.J. and Kimes, R.V. Empirical characterization of random forest variable importance measures. Computational Statistics and Data Analysis, 52, 2249- 2260 (2008).
59. Auret, L. and Aldrich, C. Empirical comparison of tree ensemble variance importance measures. *Chemometrics and Intelligent Laboratory Systems*, 105 157-170 (2011).
60. Tolosi L, Lengauer T. Classification with correlated features: unreliability of feature ranking and solutions. *Bioinformatics.* 15;27(14), 1986-94 (2011).
61. Strobl, C., Boulesteix, AL., Zeileis, A. et al. Bias in random forest variable importance measures: Illustrations, sources and a solution. *BMC Bioinformatics* 8, 25 (2007).
62. Hothorn, T., Hornik, K., Zeileis, A. Unbiased Recursive Partitioning: A Conditional Inference Framework, *Journal of Computational and Graphical Statistics*, 15:3, 651-674 (2006).
63. Xia, R. Comparison of Random Forests and Cforest: Variable Importance Measures and Prediction Accuracies. *All Graduate Plan B and other Reports.* 1255. (2009). https://digitalcommons.usu.edu/gradreports/1255.
64. Biecek, P., & Burzykowski, T. Explanatory Model Analysis: Explore, Explain, and Examine Predictive Models (1st ed.). Chapman and Hall/CRC. https://doi.org/10.1201/9780429027192.

## Acknowledgments

Aydede acknowledges financial support from Research Nova Scotia for the research project (HRC-2020-112) on Using Machine Learning to Predict Viral Transmission Rates in Halifax. Ditzen acknowledges financial support from Italian Ministry MIUR under the PRIN project Hi-Di NET - Econometric Analysis of High Dimensional Models with Network Structures in Macroeconomics and Finance (grant 2017TA7TYC).

## Author contributions

Both authors have contributed equally.

## Competing interests

The authors declare no competing interests.

## Data availability

This study uses a confidential and administrative dataset for COVID-19 tests obtained from Nova Scotia Health Authority. Its anonymized version aggregated for each FSA level is available for dominance analysis. The census data can be obtained from Census Analyzer: http://dc1.chass.utoronto.ca/census/index.html.